\documentclass{iopart}  

\usepackage{iopams}
\usepackage{amscd}
\usepackage{amstext}
\usepackage{amsopn}
\usepackage{setstack}
\usepackage{epic,eepic}
\usepackage{epsfig,subfigure}
\usepackage{subfigure}

\newcommand{\set}[1]{\ensuremath{\left\{#1\right\}}}
\newcommand{\cats}[1]{\ensuremath{\big|#1\big>}}
\newcommand{\bras}[1]{\ensuremath{\big<#1\big|}}

\begin{document}
\jl{1}
\letter{Transfer-matrix DMRG for stochastic models: \\ The Domany-Kinzel
  cellular automaton\footnote[7]{Dedicated to Prof.\ Erwin M\"uller-Hartmann on
    the occasion of his 60th birthday.}}
\author{A~Kemper\dag, A~Schadschneider\ddag\ and J~Zittartz}
\address{Institut f\"ur Theoretische Physik,
         Universit\"at zu K\"oln,
         Z\"ulpicher Str. 77, D-50937 K\"oln, Germany}   
\eads{\mailto{\dag kemper@thp.uni-koeln.de}, 
\mailto{\ddag as@thp.uni-koeln.de}}
\submitto{\JPA}


\begin{abstract}
  We apply the transfer-matrix DMRG (TMRG) to a
  stochastic model, the Domany-Kinzel cellular automaton, which
  exhibits a non-equilibrium phase transition in the directed percolation
  universality class. Estimates for the stochastic time 
  evolution, phase boundaries and critical exponents can be obtained with
  high precision. This is possible using only modest numerical effort
  since the thermodynamic limit can be taken analytically in our approach.
  We also point out further advantages of the TMRG over
  other numerical approaches, such as classical DMRG or Monte-Carlo
  simulations.
\end{abstract}


\section{Introduction}
The density-matrix-renormalisation-group (DMRG) is a powerful
numerical tool which was developed by White \cite{W92,W93} to investigate 
one-dimensional quantum chains. Since then it has been applied to
several problems in quantum and classical physics \cite{DMRG}. 

DMRG studies of one-dimensional stochastic processes represent a 
relatively new field of interest.
Such models can be described in terms of a 
non-symmetric stochastic Hamiltonian $H$ \cite{A94,S00}. The treatment of
non-symmetric matrices within a DMRG algorithm \cite{P98,H98,C99} 
is a more challenging task than that of quantum systems. 
Carlon \etal \cite{C99} have systematically studied the application of
the ``classical'' DMRG to reaction-diffusion models. They showed that
it is possible to obtain accurate results for non-equilibrium phase
transitions and their critical properties.

We propose here a different DMRG approach to study stochastic
models, namely the transfer-matrix DMRG (TMRG). This method has 
first been applied to two-dimensional classical systems by Nishino
\cite{N95}. Xiang \etal \cite{X96,X97} used the TMRG to investigate the 
quantum transfer-matrix of one-dimensional quantum systems 
and introduced an accurate method for studying thermodynamic 
properties \cite{K99}.

As an example  
we choose the Domany-Kinzel cellular automaton (DKCA) \cite{D84,K85}
which exhibits a non-equilibrium phase transition in the 
universality class of directed percolation (DP) \cite{K83,K85}. 
Its phase diagram and critical properties are well known from previous 
studies, cf.\ \cite{H00} for a review. 
We compare these literature data to our TMRG results to demonstrate 
their accuracy. 
Important advantages of our approach over DMRG and Monte-Carlo simulations 
(MCS) are pointed out, especially for systems like the DKCA for which the
infinite time limit ($t\to\infty$) and the thermodynamic limit ($N\to\infty$)
do not commute.


\section{The Domany-Kinzel cellular automaton} \label{sec:DKCA}
Cellular automata are algorithms that map one discrete configuration 
of a lattice to a new one in discrete time steps. Usually the map can
be decomposed into local update rules. In the following we
introduce the DKCA which is an example of a stochastic cellular
automaton.

\subsection{The model} \label{sec:DKCA:model}
The DKCA \cite{D84,K85} is defined on a periodically closed chain 
of $N$ sites. Each site $s_i$ can take the values $s_i\in\set{0,1}$. A 
configuration of the chain can be interpreted as a vector
$\cats{s}=\cats{s_1,\dots,s_N}$ in a "Hilbert-space". We  
speak of dead ($s_i=0$) and active ($s_i=1$) sites. 
The local update rules of a site $s_i(t+1)$ are given by
certain conditional transition probabilities
\begin{equation}
  \label{eq:prob}
   p\big(s_i(t+1)|s_{i-1}(t)s_{i+1}(t)\big)
\end{equation}
which depend only on the neighbouring sites $s_{i\pm 1}(t)$. 
The model is controlled by three independent parameters
\begin{equation}
  \label{eq:param}
  p_0:=p(1|00),\quad p_1:=p(1|01)=p(1|10) \quad \text{and} \quad
  p_2:=p(1|11)
\end{equation}
where  $p(0|\cdot)=1-p(1|\cdot)$. The DKCA is defined by $p_0=0$.
We are interested in the time evolution of an initial probability
distribution $P(t=0)$ of lattice states. The phases of the model 
are characterised by the properties of the
stationary distribution $P(t=\infty)$. 
The probability distribution $P(t)$ can be regarded as a vector   
$ \cats{P(t)}=\sum_s P_s(t) \cats{s}$
where $P_s(t)$ denotes the probability of a state $\cats{s}$ at time step
$t$. 

\subsection{Phases and Critical Phenomena} \label{sec:DKCA:phases}
The phase diagram differs, whether we consider a finite or infinite
chain length $N$. For a finite system only one phase occurs for
arbitrary $p_1,p_2<1$. Any initial state $\cats{P(0)}$ decays
exponentially fast to the stationary state 
$\cats{0}:=\cats{00\cdots 0}$
which consists of dead sites only.
$\cats{0}$ is referred to as an \emph{absorbing state} of the model
because the system cannot escape from $\cats{0}$ due to $p_0=0$. 
However, the situation changes in the thermodynamic limit
$N\to\infty$. 
For sufficiently large $p_1$ and $p_2$ an arbitrary initial state
$\cats{P(0)}\neq \cats{0}$ evolves into an unique stationary state 
$\cats{P(\infty)}\neq \cats{0}$. Thus we observe two phases, an 
\emph{absorbing} and \emph{active} one, separated by
a sharp transition which falls into the DP universality class. 

Consider an arbitrary curve $f(p)$ $(0<p<1)$ 
in the phase diagram $(p_1,p_2)$. 
An order parameter for the transition from the active to the absorbing
phase is given by the local density of active sites
\begin{equation}
  \label{eq:order1}
  n_p(t):=\bras{1} n_i \cats{P(t)}
\end{equation}
with $\cats{1}:=\sum_s \cats{s}$ and the local number operator $n_i$. 
In the absorbing phase $n_p(t)$ vanishes in the limit $t\to\infty$,
while in the active phase $n_p(t)$ saturates at some stationary 
value $n_p(\infty)$.
Close to the phase transition point $p_c$ a power law behaviour
\begin{equation}
  \label{eq:pown}
  n_p(\infty)\sim (p-p_c)^\beta
\end{equation}
is observed.
In contrast to equilibrium systems without dynamical aspects, critical 
phenomena in non-equilibrium systems are usually characterised by two
correlation lengths $\xi_\perp,\xi_\|$ in space and time respectively.
Close to a phase transition point these correlation lengths diverge as
\begin{equation}
  \label{eq:powcorr}
  \xi_\perp \sim (p-p_c)^{-\nu_\perp}, \quad
  \xi_\| \sim (p-p_c)^{-\nu_\|}.
\end{equation}
The DP universality class is already determined by the triple 
$(\beta,\nu_\perp,\nu_\|)$ \cite{H00}. 
Two other critical exponents $\alpha$ and $\sigma$ are used in this work. 
Choosing $p\approx p_c$, the dynamic evolution of $n_p(t)$ scales as 
\begin{equation}
  \label{eq:alpha}
  n_p(t)\sim t^{-\alpha},
\end{equation}
and the response of $n_p(\infty)$ to an ``external field''
$p_0$ as
\begin{equation}
  \label{eq:sigma}
  n_{p,p_0}(\infty)\sim (p_0)^{\beta / \sigma}.
\end{equation}
$\alpha$ and $\sigma$ can be related to 
$\beta,\nu_\perp$ and $\nu_\|$ using scaling theory \cite{H00}:
\begin{equation}
  \label{eq:scaling}
  \alpha=\frac{\beta}{\nu_\|} \quad \mathrm{and} \quad
  \sigma=\nu_\| + \nu_\perp - \beta.
\end{equation}


\section{Transfer-matrix DMRG} \label{sec:TMRG}
We first map the DKCA onto a two-dimensional classical system which 
can be solved using a transfer-matrix formulation. 
In the thermodynamic limit $N\to\infty$ the properties of the model
can be inferred from the knowledge of certain eigenvalues and -vectors 
of the transfer-matrix.
The TMRG algorithm calculates this dominating part of the spectrum.

\subsection{Mapping} \label{sec:TMRG:mapping}
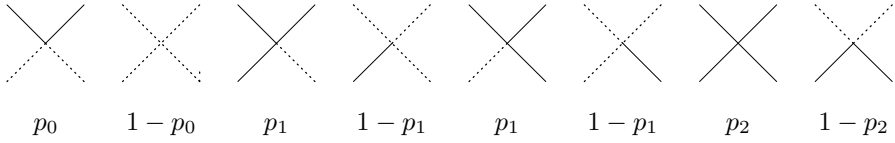
\begin{figure}[ht]
  \begin{center}
    {\setlength{\unitlength}{0.00066667in}
{\renewcommand{\dashlinestretch}{30}
\begin{picture}(6924,1051)(0,-10)
\path(312,724)(12,1024)
\path(312,724)(612,1024)
\path(1812,424)(2112,724)
\path(2112,724)(1812,1024)
\path(2112,724)(2412,1024)
\path(2712,424)(3012,724)
\path(4212,424)(3912,724)
\path(3912,724)(3612,1024)
\path(3912,724)(4212,1024)
\path(5112,424)(4812,724)
\path(5412,424)(6012,1024)
\path(6012,424)(5412,1024)
\path(6312,424)(6612,724)
\path(6612,724)(6912,424)
\dottedline{45}(12,424)(312,724)
\dottedline{45}(312,724)(612,424)
\dottedline{45}(912,424)(1512,1024)
\dottedline{45}(2112,724)(2412,424)
\dottedline{45}(2712,1024)(3312,424)
\dottedline{45}(3012,724)(3312,1024)
\dottedline{45}(3612,424)(3912,724)
\dottedline{45}(4512,424)(5112,1024)
\dottedline{45}(4512,1024)(4812,724)
\dottedline{45}(6312,1024)(6612,724)
\dottedline{45}(6612,724)(6912,1024)
\dottedline{45}(912,1024)(1512,424)(1512,499)
\put(312,49){\makebox(0,0)[b]{\smash{$p_0$}}}
\put(1212,49){\makebox(0,0)[b]{\smash{$1-p_0$}}}
\put(2112,49){\makebox(0,0)[b]{\smash{$p_1$}}}
\put(3012,49){\makebox(0,0)[b]{\smash{$1-p_1$}}}
\put(3912,49){\makebox(0,0)[b]{\smash{$p_1$}}}
\put(4812,49){\makebox(0,0)[b]{\smash{$1-p_1$}}}
\put(5712,49){\makebox(0,0)[b]{\smash{$p_2$}}}
\put(6612,49){\makebox(0,0)[b]{\smash{$1-p_2$}}}
\end{picture}
}}
    \caption{Vertices of the classical model with 
      corresponding probabilities. Full lines correspond to bonds.}
    \label{fig:vertices}
  \end{center}
\end{figure}

The local update probabilities \eref{eq:prob} of a 
site $s_i(t+1)$ only depend on the neighbouring sites 
$s_{i\pm 1}(t)$. Hence, we 
use a sub-lattice (parallel) update. An update time step $t\to t+1$
is divided into two sub-steps updating the sub-lattice of odd and even sites 
separately. 
\begin{figure}[ht]
  \begin{center}
    \input{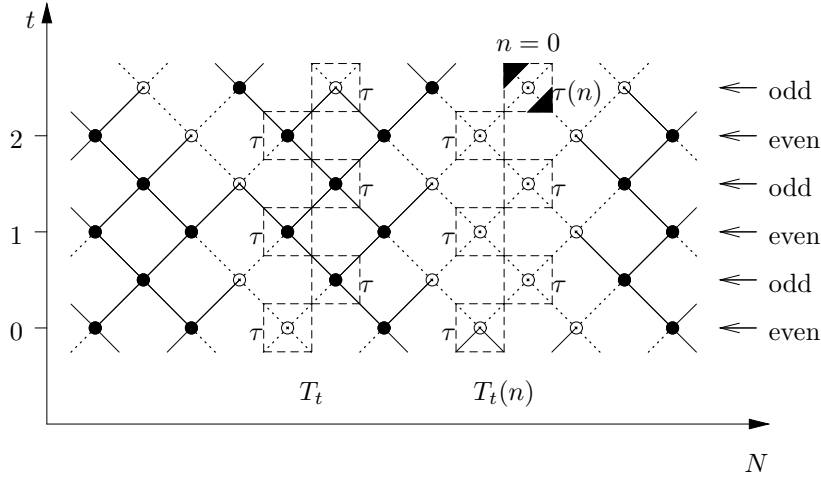}
    \caption{Sub-lattice parallel update of the DKCA.}
    \label{fig:spu}
  \end{center}
\end{figure}
\Fref{fig:spu} depicts a realization of the update. The sites are 
connected by two state bonds. A bond is present (full line), if the 
predecessor  $s_{i\pm 1}(t)$ of a site $s_i(t+1)$ is active, or not
present (dotted line) otherwise. As shown in \fref{fig:spu}, the bonds form a 
two-dimensional square lattice $\Gamma$. A configuration $\gamma \in
\Gamma$ of the lattice decomposes into vertices $\nu$ which can be in
eight different states. These are shown in \fref{fig:vertices}
together with the probabilities assigned to them.
 
Various properties of the stochastic system can now be expressed in
terms of the statistical mechanics of a two-dimensional vertex model. 
For determining the critical behaviour we are mainly interested in 
the order parameter 
\begin{equation}
  \label{eq:order2}
  n_p(t)=\frac{\sum_{\gamma\in\Gamma} n(\gamma) p(\gamma)}
  {\sum_{\gamma\in\Gamma} p(\gamma)}.
\end{equation}
$p(\gamma)=\prod_{\nu\in\gamma} \nu$ denotes the probability of a 
certain lattice configuration $\gamma$ and corresponds to the
statistical weight in a vertex
model. $n(\gamma)\in\set{0,1}$ measures
the state of a certain lattice site, cf.\ \fref{fig:spu}. Thus
the calculation of $n_p(t)$ requires taking the average 
$\left<n\right>$ of a local operator $n$ in a vertex model.  
Note that we have periodic boundary conditions in 
space direction $N$, but open ones in time direction $t$.  

\subsection{Transfer-matrix formulation}
We briefly recall some basic facts of the transfer-matrix
formulation of two-dimensional classical systems. The idea is
to transform the sum over all configurations in
\eref{eq:order2} into a product of transfer-matrices. 
The vertex model representing the dynamics
of the DKCA has a checkerboard-like structure, cf.\ \fref{fig:spu}. 
Each dashed square tags a vertex and defines a local transfer matrix
\begin{equation}
  \label{eq:tau}
  \fl
  \tau = \left(\bras{\sigma_1\sigma_2} \tau
  \cats{\sigma_1'\sigma_2'}\right) = 
  {\setlength{\unitlength}{0.00083333in}
\renewcommand{\dashlinestretch}{30}
\begin{picture}(800,400)(600,200)
      \dashline{60.000}(847,433)(1147,433)(1147,133)
      (847,133)(847,433)
      \path(847,433)(1147,133)
      \path(847,133)(1147,433)
      \put(847,58){\makebox(0,0)[rb]{\smash{{{$\sigma_1$}}}}}
      \put(847,358){\makebox(0,0)[rb]{\smash{{{$\sigma_2$}}}}}
      \put(1147,358){\makebox(0,0)[lb]{\smash{{{$\sigma_2'$}}}}}
      \put(1147,58){\makebox(0,0)[lb]{\smash{{{$\sigma_1'$}}}}}
\end{picture}}=
  \left(
  \begin{array}{cccc}
    1-p_0&0&1-p_1&0\\
    0&p_0&0&p_1\\
    1-p_1&0&1-p_2&0\\
    0&p_1&0&p_2
  \end{array}
  \right).
\end{equation}
The (vertical) transfer-matrix $T_t$ of a certain time step $t$ 
is defined by the matrix elements
\begin{equation}
  \label{eq:transfer}
  \fl
  \bras{\sigma_2\cdots\sigma_{2t}} T_t \cats{\sigma_2'\cdots\sigma_{2t}'}
  =\hspace{-8pt}\sum_{\substack{\set{\tilde\sigma_k}\\\sigma_1\sigma_{2t+1}'}} 
  \hspace{-5pt}\prod_{k=1}^t
  \bras{\sigma_{2k-1}\sigma_{2k}}\tau
  \cats{\tilde\sigma_{2k-1}\tilde\sigma_{2k}}
  \bras{\tilde \sigma_{2k}\tilde \sigma_{2k+1}}\tau
  \cats{\sigma_{2k}'\sigma_{2k+1}'}.
\end{equation}
In \eref{eq:transfer} the open boundary conditions in
$t$-direction are considered by summing out $\sigma_1$, $\tilde
\sigma_1$, $\sigma_{2t+1}'$ and $\tilde \sigma_{2t+1}$.
Note that this definition is similar to the so-called quantum transfer
matrix used in \cite{X96,X97,K99} for quantum systems. In contrast to
the horizontal transfer matrix in $N$-direction, $T_t$ is not a
stochastic matrix. 

For the calculation of the order parameter another matrix $T_t(n)$ is
needed which contains one modified local transfer matrix (see
\fref{fig:spu})
\begin{equation}
  \label{eq:taun}
  \tau(n)= \left(
  \begin{array}{cccc}
    0&0&0&0\\
    0&p_0&0&p_1\\
    0&0&0&0\\
    0&p_1&0&p_2
  \end{array}
  \right).
\end{equation}
Using $T_t$ and $T_t(n)$ it can be easily shown that
\begin{equation}
  \label{eq:ntrans}
  n_p(t)=\frac{\tr\left(T_t(n)
  T_t^{N/2-1}\right)}{\tr\left(T_t^{N/2}\right)}.
\end{equation}
Under the assumption that the highest
eigenvalue $\Lambda_0$ of $T_t$ is not degenerate, 
the thermodynamic limit $N\to\infty$ can be
performed exactly
\begin{equation}
  \begin{CD}
  \label{eq:limit}
  n_p(t)={\displaystyle \frac{\tr\left(T_t(n)
  T_t^{N/2-1}\right)}{\tr\left(T_t^{N/2}\right)}}
  @>N\to\infty>>
  {\displaystyle
    \frac{\bras{\Lambda_0^L}T_t(n)\cats{\Lambda_0^R}}{\Lambda_0}}. 
  \end{CD}
\end{equation}
$\cats{\Lambda_0^R}$ ($\cats{\Lambda_0^L}$) denotes the right (left)
eigenvector of the non-symmetric transfer matrix $T_t$ corresponding
to the eigenvalue $\Lambda_0$. 

\subsection{TMRG algorithm}
The TMRG calculates $\Lambda_0$ and the corresponding left and
right eigenvector for an arbitrary matrix size $t$. This is done
iteratively by numerical renormalisation techniques.   
We do not go into the details of the algorithm,
which is basically the same used for the quantum transfer-matrix in
\cite{X97}. Two important modifications are necessary:
\begin{itemize}
\item 
  Open boundary conditions in time direction
  $t$ are used in \eref{eq:transfer}. 
  These are necessary as the DKCA is not 
  translationally invariant in time. In contrast, the corresponding Trotter
  dimension in quantum-TMRG is periodically closed \cite{X97}.
\item 
  The density matrix   
  $\rho_{\mathrm{ns}}=\tr_{\mathrm{e}} 
  \cats{\Lambda_0^R} \bras{\Lambda_0^L}$
  used in quantum-TMRG 
  is not symmetric where $\tr_{\mathrm{e}}$ denotes the partial trace
  over the environmental block. 
  Using $\rho_{\mathrm{ns}}$ for the DKCA, we observed a poor
  convergence of the TMRG algorithm.
  Following \cite{C99} we restricted ourselves to the symmetric density matrix 
  \begin{equation}
    \label{density}
    \rho_{\mathrm{s}} = \frac{1}{2} \tr_{\mathrm{e}} \left( \cats{\Lambda_0^R}
      \bras{\Lambda_0^R} + \cats{\Lambda_0^L} \bras{\Lambda_0^L}
      \right).
  \end{equation}
  The convergence of the algorithm
  was improved remarkably.
\end{itemize}
The eigenvectors $\cats{\Lambda_0^R}$ and $\cats{\Lambda_0^L}$ 
of $T_t$ were calculated by a power method. The diagonalisation of
the density matrix was done by using \textsc{Maple} which can in
principle handle arbitrary numerical precision. The data presented in the
following section have been obtained by keeping $m=24$ states. 


\section{Results} \label{sec:res}
\subsection{Phase diagram}
In the absorbing phase $n_p(t)$ decreases exponentially fast to zero,
whereas $n_p(\infty)\neq 0$ in the active phase. At the transition point
$p=p_c$ we expect a power law behaviour. 
\begin{figure}[ht]
  \begin{center}
    \subfigure{\raisebox{4.5cm}{$\ln n_p(t)$}
      \hspace{-0.6cm}
      \epsfig{figure=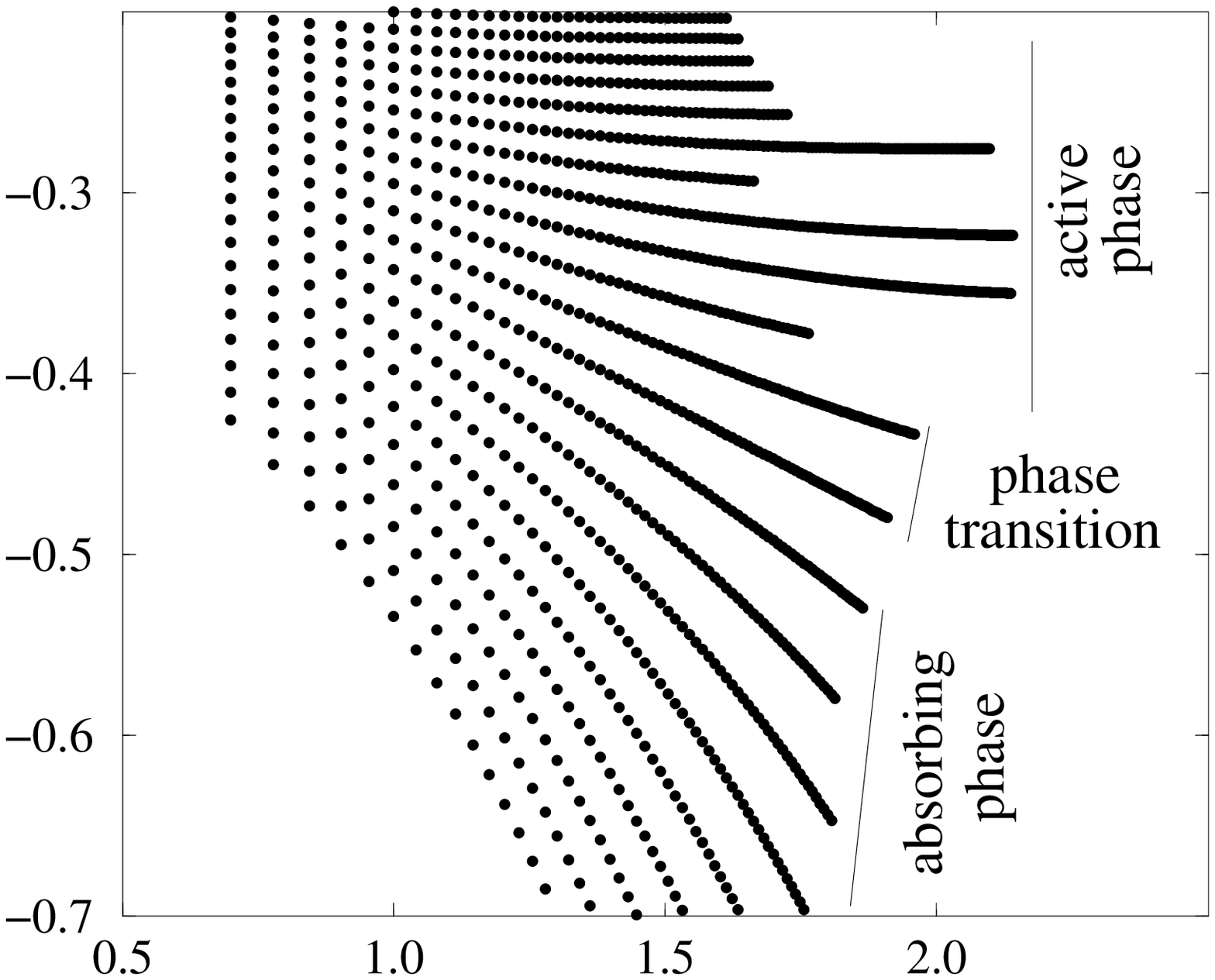,width=.45\textwidth}
      \hspace{-0.6cm}\raisebox{-1mm}{$\ln t$}}\hspace{0.3cm}
    \subfigure{\raisebox{4.6cm}{$p_2$}
      \hspace{-0.5cm}
      \epsfig{figure=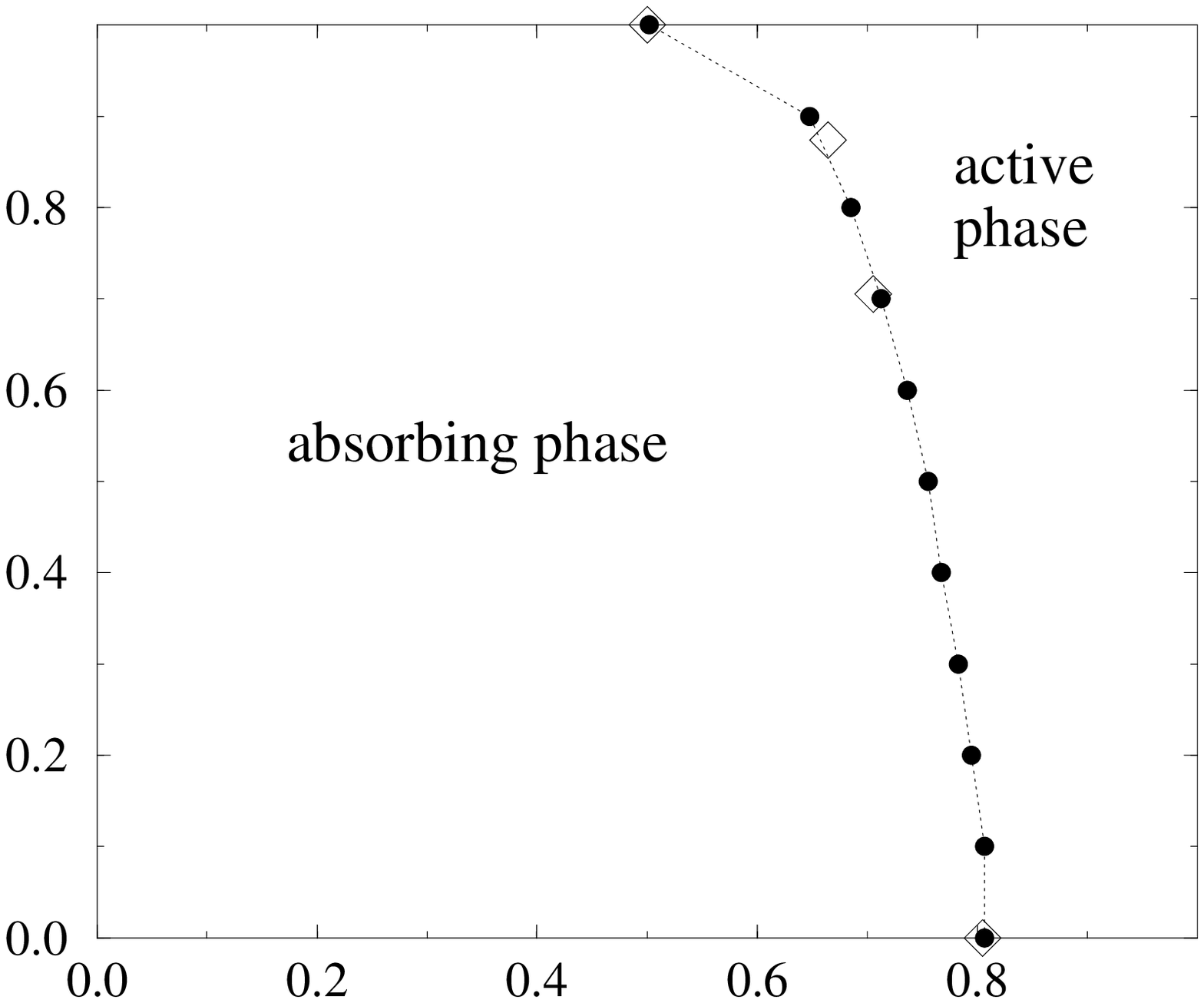,width=.45\textwidth}
      \hspace{-0.5cm}\raisebox{-1mm}{$p_1$}} \vspace{-.5cm}
    \caption{The left figure shows the dynamic evolution of $n_p(t)$ for
      different $p$ on the SDP line. The curvature of the log/log-plot 
      determines whether $p$ belongs to the active or absorbing
      phase. The right figure depicts the phase diagram  
      calculated using TMRG ($\bullet$) in comparison with data ($\diamond$) 
      from \cite{K83,R94,T95,J96}.}
    \label{fig:phases}
  \end{center}
\end{figure}
\Fref{fig:phases} (left) presents a
double-logarithmic plot of $n_p$ along a curve $f(p)$
in the phase diagram $(p_1,p_2)$. As an example we used the
site-directed percolation (SDP) line $f_{\text{SDP}}(p)=(p,p)$.
$p_c$ is determined by adjusting $p$ until the curvature of $n_p(t)$
in the double-logarithmic plot vanishes. This was
done for various lines $f_{p_2}(p)=(p,p_2)$ to calculate the complete
transition line, cf.\ \fref{fig:phases} (right).

\subsection{Order parameter}
In a second step we estimated $n_p(\infty)$ in the
active phase by a $1/t$ finite-size scaling. 
\begin{figure}[ht]
  \begin{center}
    \raisebox{6.7cm}{$n_p(\infty)$}
    \hspace{-.7cm}
    \epsfig{file=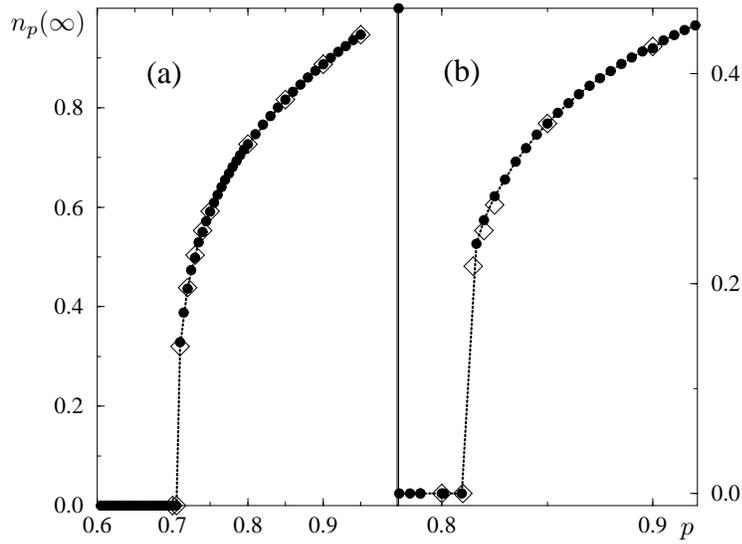,width=0.7\textwidth}
    \hspace{-.9cm}\raisebox{0mm}{$p$}
    \caption{TMRG results ($\bullet$) of $n_p(\infty)$ 
      obtained from calculations across (a) the SDP-line
      and (b) the line $f_0(p)=(p,0)$. MCS data ($\diamond$) are 
      plotted for comparison. }
    \label{fig:order}
  \end{center}
\end{figure}
\Fref{fig:order} depicts the results we obtained
for the SDP line and the line $f_0(p)=(p,0)$. 
Within the phases our data are in excellent agreement with MCS. 
Only close to the critical point we observe a small 
deviation.
Here the $1/t$-scaling becomes less accurate due to 
critical slowing down. 
We expect that other methods, i.e.\ VBS or BST
\cite{B64,H88}, provide better results.   

\subsection{Critical exponents}
The critical exponents $\alpha$, $\beta$ and $\sigma$ determine the 
universality class, cf.\ \sref{sec:DKCA:phases}.
We exemplify the calculation of the exponents by investigating
the SDP line. 
\begin{figure}[ht]
  \begin{center}
    \leavevmode
    \subfigure{\raisebox{4.5cm}{$n_p(t)$}
      \hspace{-0.6cm}
      \epsfig{file=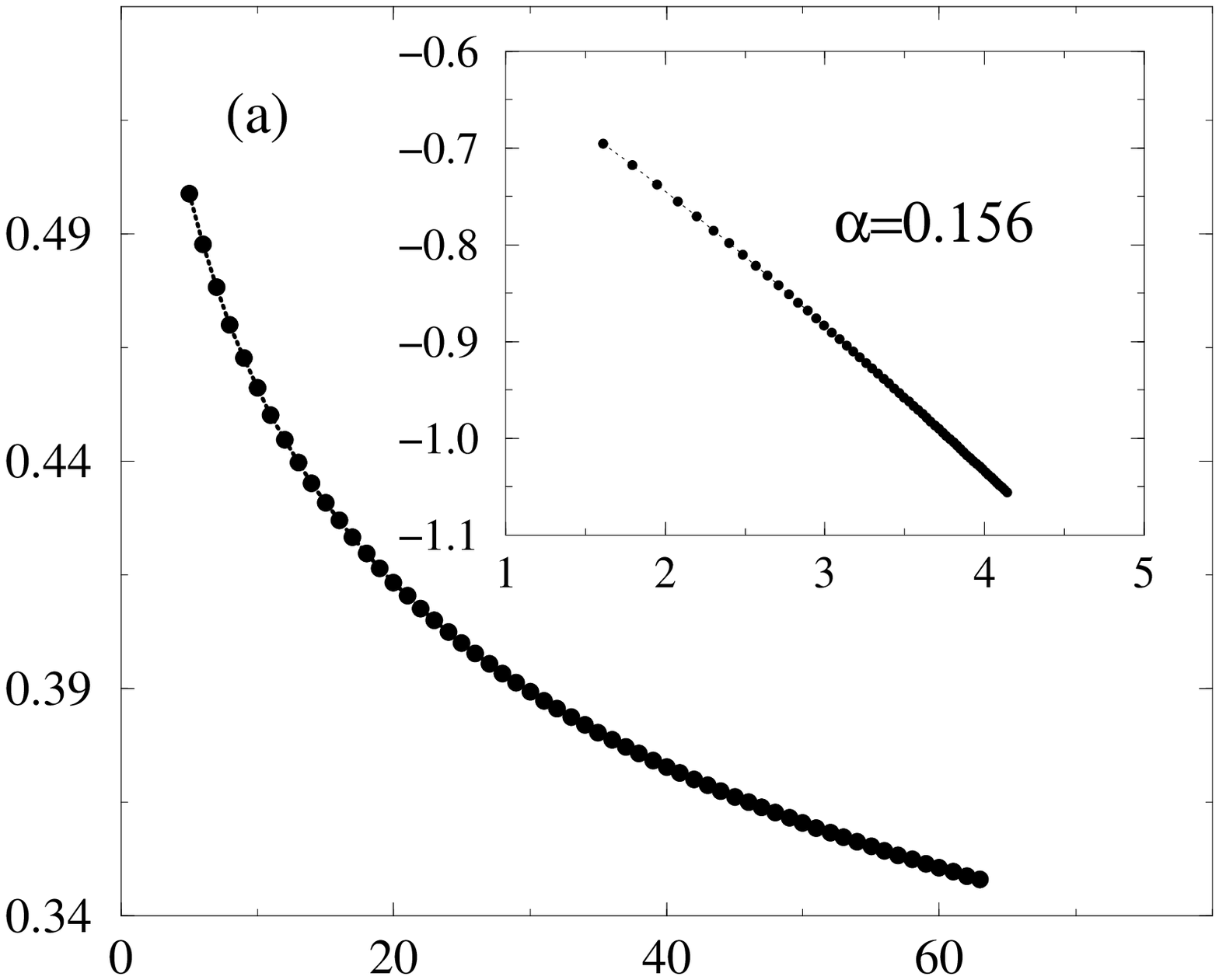,width=0.45\textwidth}
      \hspace{-0.4cm}\raisebox{-1mm}{$t$}
      }\hspace{0.2cm}
    \subfigure{\raisebox{4.6cm}{$n_p(\infty)$}
      \hspace{-0.5cm}
      \epsfig{file=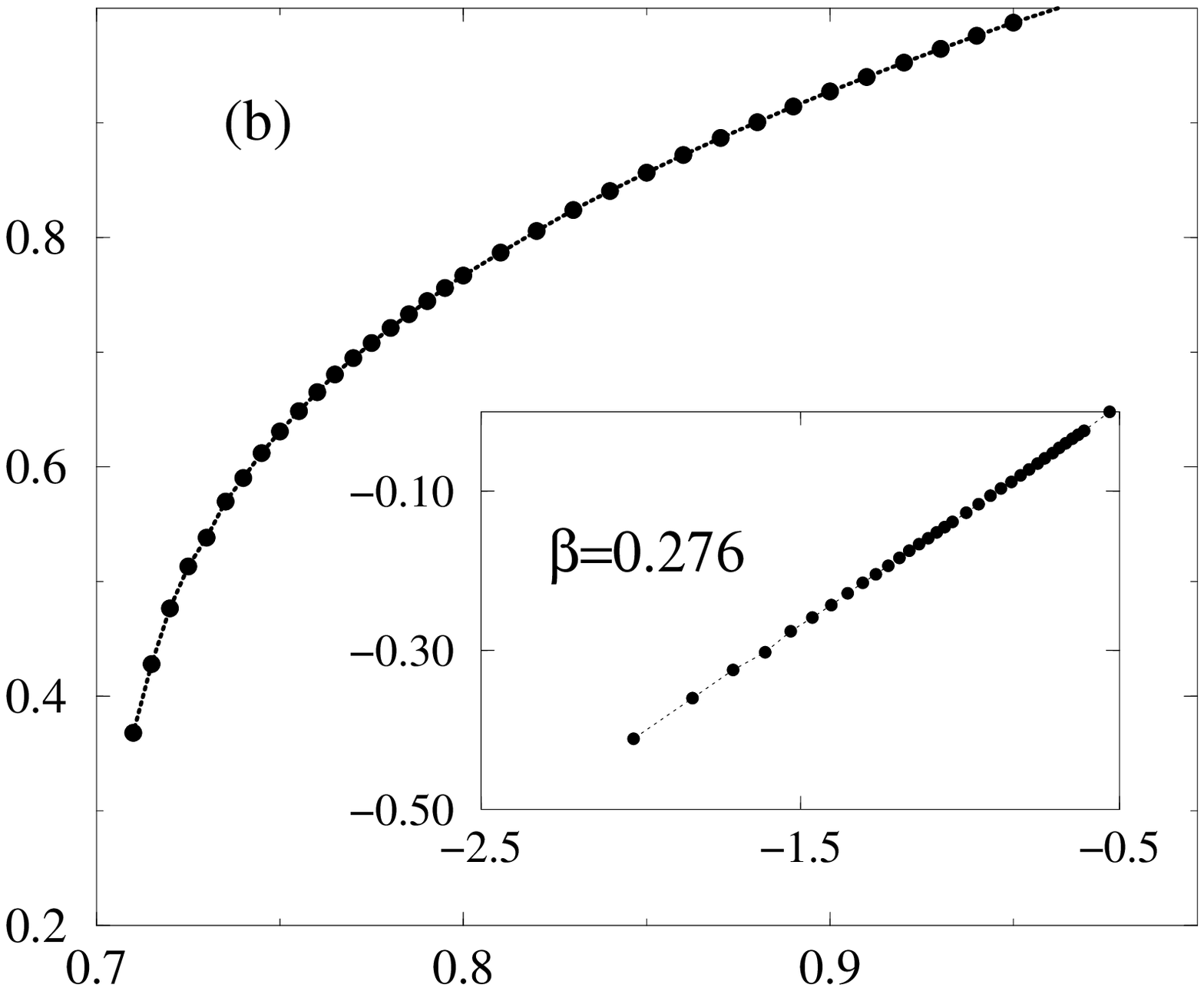,width=0.45\textwidth}
      \hspace{-0.4cm}\raisebox{-1mm}{$p$}} 
    \subfigure{
      \raisebox{4.5cm}{$n_{p,p_0}(\infty)$} \hspace{-0.6cm}
      \epsfig{file=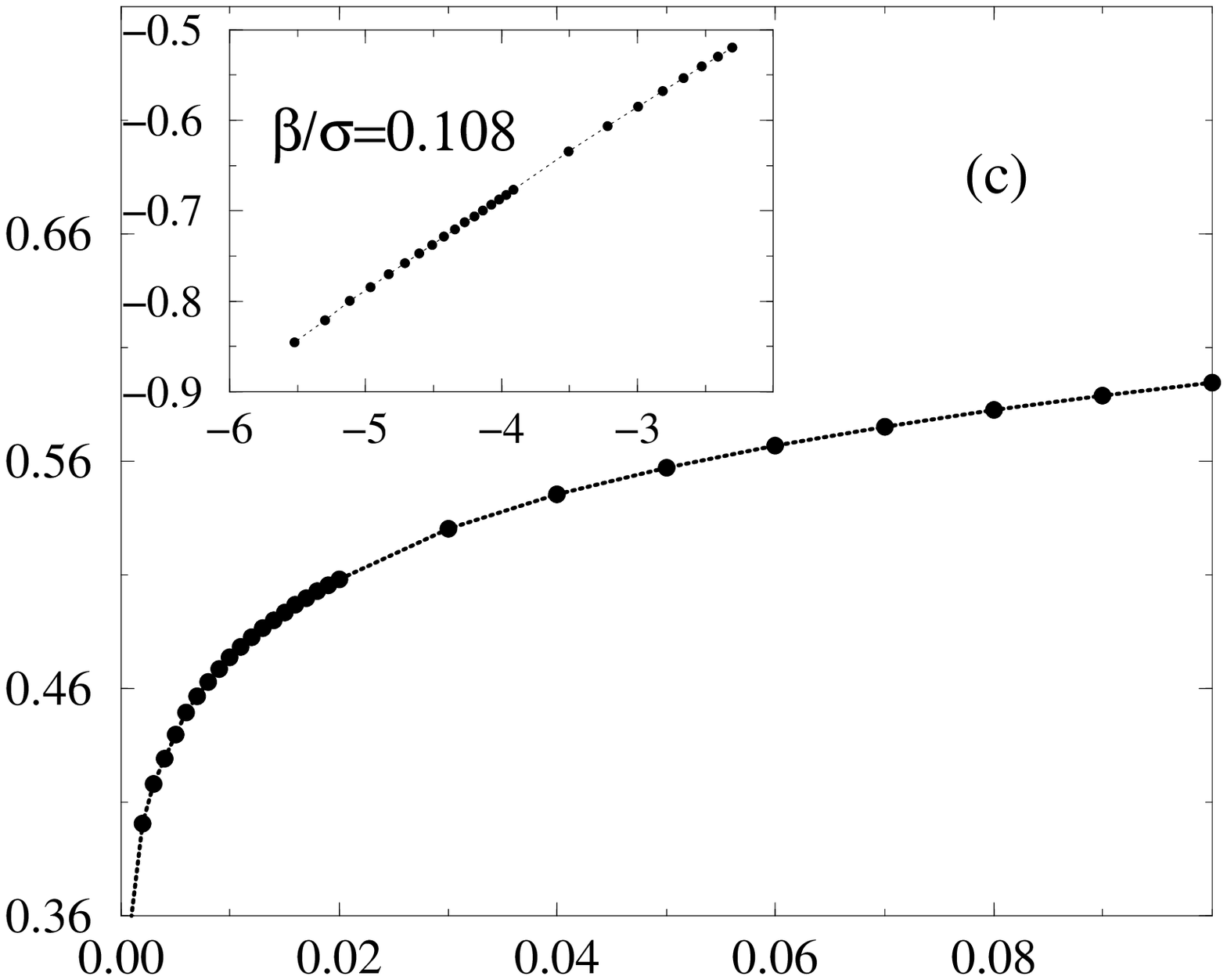,width=0.45\textwidth}
      \hspace{-0.4cm}\raisebox{-1mm}{$p_0$}}
    \caption{Calculation of the critical exponents (a) $\alpha$,
    (b) $\beta$ and (c) $\frac{\beta}{\sigma}$. The insets are 
    double-logarithmic plots of the particular data. The critical exponents 
    are calculated by determining the linear regression slope of 
    the log/log-plot. }
    \label{fig:crit}
  \end{center}
\end{figure}
\Fref{fig:crit} depicts for the order parameter (a) its time-dependence
$n_p(t)$, (b) the stationary value $n_p(\infty)$ and 
(c) the stationary value $n_{p,p_0}(\infty)$ in the presence of a ``field''
$p_0\neq 0$ with $p$ close to $p_c$.    
The linear regression slope of a double-logarithmic plot (see insets)
yields the particular exponent. 
\Tref{tab:res} compares the results with other data obtained by series
expansion and MCS.
The accuracy of the exponents
is similar to the DMRG calculations done by Carlon \etal \cite{C99}.   
\begin{table}
  \lineup
  \begin{indented}
    \item[]\begin{tabular}{@{}cccc}
    \br
    critical      &series     &MCS        &TMRG\\
    exponent      &\cite{J99} &\cite{L97,M98}    &(this work)\\
    \mr
    $\beta$       &0.276486(8)\0 &0.27649(4)&0.276(5)\\
    $\nu_\|$      &1.733847(6)\0 &1.73383(3)&\\
    $\nu_\perp$   &1.096854(4)\0 &1.09684(1)&\\
    $\alpha$      &0.159464(5)\0 &0.15946(2)&0.156(5)\\    
    $\beta/\sigma$&0.10825(2)\0\0&0.10825(2)&0.108(5)\\
    \br
    \end{tabular}
    \caption{Estimates for the critical exponents obtained by
    series expansion and MCS in comparison with TMRG. Results for
    $\nu_\|$ and $\nu_\perp$ can be obtained for our data using the
    scaling relations (\ref{eq:scaling}).}
    \label{tab:res}
  \end{indented}
\end{table}


\section{Conclusions and Outlook} \label{sec:concl}
In this letter we have proposed a new approach to
stochastic models using TMRG. The accuracy of the results were shown by
comparison to MCS or literature data. Critical exponents were
estimated which deviate from the currently accepted values \cite{J99}
by less than 3\%.
In the following we focus on three important advantages
of TMRG compared to traditional MCS or DMRG calculations. 

First, the exact treatment of the thermodynamic limit of
the chain is the main aspect of the TMRG presented here. 
In contrast, MCS and DMRG can only be
applied to finite chains. Thus a finite-size scaling has to be used
carefully to estimate the thermodynamic limit $N\to\infty$. For transitions
into absorbing states $N\to\infty$ does usually 
not commute with the infinite time limit $t\to\infty$. 
As shown in \eref{eq:limit}, $N\to\infty$ can be performed
exactly within the transfer-matrix formulation.

Second, the TMRG approach facilitates a proper investigation of the
dynamic evolution of the system (i.e.\ $n_p(t)$). This is a more difficult
task using MCS. Many samples have to be taken into account to obtain a
reasonable average. 
In contrast, the DMRG and TMRG approaches do not require random numbers
and averaging like the MCS.
The DMRG algorithm basically calculates the ground
state of the stochastic Hamiltonian $H$ which corresponds to the steady
state ($t\to\infty$) \cite{C99}. Dynamic properties are available by analysing
the gap of $H$. Thus the second eigenvalue of $H$ has to be computed.
Critical properties are then extracted from the finite-size data of the
gap \cite{C99}.
On the other hand, the TMRG offers direct access to the dynamics and phase
transition of the model by calculating one eigenvalue only.

The third advantage is a rather technical one. We obtained
surprisingly accurate results by modest numerical efforts. 
In comparison, the classical DMRG algorithm needs more sophisticated
numerical techniques to guarantee a sufficient convergence \cite{C99}.
We expect a further improvement
of our results by involving various numerical refinements, i.e.\ a
modified Arnoldi method \cite{A00} or a BST finite size scaling 
\cite{B64,H88}.

The TMRG approach can be generalized to study other stochastic processes
with arbitrary dynamics.
Currently we investigate the pair contact process with diffusion
(PCPD) and compare TMRG and DMRG results \cite{H01}.
The universality class of the model is
still unknown and an interesting field of research \cite{C00,H00.2}. 

\ack

The authors have benefited from intensive discussions with R.~Raupach, who 
placed his TMRG program for quantum systems to our disposal. 
We also thank M.~Henkel and U.~Schollw\"ock for very interesting suggestions
and discussions about numerical aspects.
A.K. thanks J.~Sirker for very useful discussions about the TMRG 
algorithm. This work was supported by SFB 341.



\section*{References}
\begin{thebibliography}{99}
\bibitem{W92} White S R 1992 \PRL {\bf 69} 2863
\bibitem{W93} White S R 1993 \PR B {\bf 48} 10345
\bibitem{DMRG} Peschel I, Wang X, Kaulke M and Hallberg K (eds.) 1998 
  {\it Lect.~Not.~Phys.}  vol 528 (Berlin: Springer)
\bibitem{A94} Alcaraz F C, Droz M, Henkel M and Rittenberg V 1994
  \APNY {\bf 230} 250 
\bibitem{S00} Sch\"utz G 2000 in {\it Phase Transitions
and Critical Phenomena} Domb C and Lebowitz J (eds.) vol 19
(London: Academic Press)
\bibitem{P98} Kaulke M and Peschel I 1998 {\it Eur.~Phys.~J.} B {\bf 5} 727 
\bibitem{H98} Hieida Y 1998 \JPSJ {\bf 67} 369
\bibitem{C99} Carlon E, Henkel M and Schollw\"ock U 1999 
  {\it Eur.~Phys.~J.} B {\bf 12} 99
\bibitem{N95} Nishino T 1995 \JPSJ {\bf 64} 3598
\bibitem{X96} Bursill R J, Xiang T and Gehring G A 1996 \JPCM {\bf 8}
  L583
\bibitem{X97} Wang X and Xiang T 1997 \PR B {\bf 56} 5061
\bibitem{K99} Kl\"umper A, Raupach R and Sch\"onfeld F 1999 \PR B {\bf
    59} 3612
\bibitem{D84} Domany E and Kinzel W 1984 \PRL {\bf 53} 311
\bibitem{K85} Kinzel W 1985 \ZP {\bf 58} 229
\bibitem{K83} Kinzel W 1983 {\it Ann.~Isr.~Phys.~Soc.} vol~5
  (Bristol: Adam Hilger)
\bibitem{H00} Hinrichsen H 2000 {\it Adv.~Phys.} {\bf 49} 815
\bibitem{R94} Rieger H, Schadschneider A and Schreckenberg M 1994 \JPA
  {\bf 27} L423
\bibitem{T95} Tretyakov A Y and Inui N 1995 \JPA {\bf 28} 3985
\bibitem{J96} Jensen I 1996 \PRL {\bf 77} 4988
\bibitem{B64} Burlisch R and Stoer J 1964 {\it Numer.~Math.} {\bf 6}
  413
\bibitem{H88} Henkel M and Sch\"utz G 1988 \JPA {\bf 21} 2617
\bibitem{J99} Jensen I 1999 \JPA {\bf 32} 5233
\bibitem{L97} Lauritsen K B, Sneppen K, Markosov\'a M and Jensen M H
  1997 {\it Physica} A {\bf 246} 1
\bibitem{M98} Mun\~os M A, Dickmann R, Vespignani A and Zapperi S 
  1999 \PR E {\bf 59} 6175
\bibitem{A00} Henkel M and Schollw\"ock U 2000 private communications
\bibitem{H01} Henkel M, Kemper A, Schadschneider A, Schollw\"ock U and
  Zittartz J 2001, work in progress
\bibitem{C00} Carlon E, Henkel M and Schollw\"ock U 2001 \PR E {\bf
    63} 036101
\bibitem{H00.2} Hinrichsen H 2001 \PR E {\bf 63} 036102
\end{thebibliography}
\end{document}